\documentstyle[prb,preprint,aps]{revtex}
\tightenlines

\def\lsim{\mathrel{\rlap{\lower4pt\hbox{\hskip1pt$\sim$}}
    \raise1pt\hbox{$<$}}}         
\def\gsim{\mathrel{\rlap{\lower4pt\hbox{\hskip1pt$\sim$}}
    \raise1pt\hbox{$>$}}}         
\def\be{\begin{equation}}
\def\ee{\end{equation}}
\def\bq{\begin{eqnarray}}
\def\eq{\end{eqnarray}}
\def\bqn{\begin{eqnarray*}}
\def\eqn{\end{eqnarray*}}

\begin{document}

\draft
\title{Testing the helix model for protein folding on four simple proteins}
\author{Pierpaolo  Bruscolini }
\address{Unit\`a INFM, Politecnico di Torino, \\
C.so Duca degli Abruzzi 24, 10129 Torino, ITALY}

\date{\today}
\maketitle
\begin{abstract}
We test a simplified, local version of the helix model~\cite{brus97} on
two synthetic and two natural proteins, to study its efficiency in
predicting the native secondary structure. The results we obtain are
very good for the synthetic sequences, poorer for the two
natural ones. This suggests that non-local terms  play 
a fundamental role
in determining the secondary structure, even if in some cases local
terms alone may be sufficient.
\end{abstract}

\section{Introduction}
It is experimentally known~\cite{anfi73,anfi75,chot92} that a
protein, under proper solvent and temperature conditions,
folds from any random-shaped  state to its ``native''
state, whose three dimensional structure is unambiguously encoded in
the amino acid sequence. This state is the only one in which the protein
is biologically active, and is strictly related to the chemical function
of the protein. Unfortunately, experimental determination of this state is
usually a rather difficult task, and it would be highly desirable to
know how to predict the structure just from the sequence.

In spite of  many years of efforts in the field,  
a clear understanding of protein folding has not been achieved yet. 
The best results
in structure prediction are obtained by algorithms which compare the
protein under study with a database of sequences of already known structure.
This approach, even when successful, does not shed any light on the
underlying physics.

Because of the wide range of time scales involved
($10^{-13} \div 10^2 s$ and to the complexity of the system, {\sl ab-initio}
simulations of the folding process are today (and will most probably be for
a long time)  out of
reach~\cite{kash92}.  They could help to understand the fast events involved 
in the folding process.

Simple simplified physical models have been proposed 
to capture the most relevant
aspects of the problem. In these approaches the protein is 
usually described as a
chain endowed with ``charged'' beads, representing the residues, which attract
each other according to their nature. These model have been studied both
on- and 
off-lattice~\cite{sasha94,klth96,shfaa91,soon95,brona95,onwoa95,irpea97}), 
resorting to Monte Carlo simulations.

Most of the theoretical understanding of the thermodynamics and  dynamics 
of the folding process comes form these models; yet their relationship
with natural proteins is somewhat qualitative, since there is no 
well defined mapping between the configuration spaces of real and
model proteins. Hence, it is difficult to say which results can be
extended to real proteins
and which are model dependent, and the debate is still 
opened~\cite{thkl96,klth96b,tibra97} on the
identification of the relevant features that distinguish a good folder
from  a poor one. 
 
In a recent paper~\cite{brus97} we proposed a new model, which gives a
coarse-grained description of a protein in terms of helices. This
choice stems from  the fact that the elements
of native secondary structure can be well approximated resorting to
one or a few helices. Even loops can be partitioned in smaller parts and
approximated in such a way: of course their description will not be
as good as that of $\alpha$ helices. However, loops are
usually found at the surface of the native globule and are affected by
less severe geometrical constraints, so a less precise representation
should not be a major problem.

The motivation  for a coarse-grained description is related to the
fact that a certain degree of redundancy is observed in structure
encoding: several different sequences are known to fold to essentially
the same structure. This suggests that an appropriate average 
description of the sequence could be enough to predict most of the
features of the native states (even if the details of the three
dimensional structure are probably related to close packing of
side-chain, and cannot be easily captured in a simplified description).
  
In this letter we use a simplified version of the helix model,
where nonlocal interactions are neglected, to predict the secondary
structure of two synthetic~\cite{kasca93} and two natural proteins,
which are known to fold into a ``four-helix bundle''. We aim on the
one hand to
test the reliability of our model, and on the other to understand the
role of local periodicities of polar and non-polar
residues in determining the secondary structure of the protein.

The paper is organized as follows: in Sec.~\ref{sec:mod} we recall the
main characteristics of the model, in Sec.~\ref{sec:an} 
we test the efficiency of the local version of the model in
predicting the secondary structure of the four proteins;
finally, in Sec.~\ref{sec:conc}, we briefly summarize and comment our results.

\section{The model}
\label{sec:mod}

Considering that their secondary structure is a fairly general feature of
native states, and that its elements can be well represented by regular
helices, we describe any protein configuration as a continuous
curve made of pieces of helices sequentially linked together.
This description is particularly suited for
$\alpha$ and $3_{10}$ helices, but it can also be applied to
$\beta$-strands (which, in the ideal case, are helices with two residues
per turn) and, to a lesser extent, to the finite class of
tight-turns presently known and to coil regions, once they are divided
into smaller parts.

The equation of the curve representing the protein chain is assumed to be:
\begin{equation}
{\bf r}(s) \,=\, \sum_{i=1}^{N_h} b_i(s)\,{\bf h}_i(s) \;\;\; ,
\label{chain}
\end{equation}
where the parameter $s$ ranges from $0$ to $N$, the total number of residues;
$b_i(s)=1$ if $s \in
)s_{i-1},s_i($ and $b_i(s)=1/2$ if $s = s_{i-1}$ or $s=  s_i $ whereas  
$b_i(s)=0$ if $s \notin (s_{i-1},s_i)$.    

The ${\bf h}_i$ are the helices expressed in their reference frame
(${\bf e}_{1,i},{\bf e}_{2,i},{\bf e}_{3,i}$):
\begin{eqnarray}
{\bf h}_i(s) &=& a_i \,\left[\, \left( \cos(u_i(s-s_{i-1}))-1
\right)\, {\bf e}_{1,i} \,+ 
\sin(u_i(s-s_{i-1})) \,{\bf e}_{2,i} \,+ \right. \nonumber \\ 
&& \left.  u_i h_i(s-s_{i-1})\,{\bf e}_{3,i}\,\right] + 
{\bf h}_{i-1}(s_{i-1}) \;\;\;,
\label{hel}
\end{eqnarray}
labelled so that helix $i$ starts at $s_{i-1}$ and ends at
$s_i$, with $s_0=0$ and $s_{N_h}= N$. $N_h$ is the total number of
helices, residues are labeled from 1 to $N$, and the convention holds
that a residue sitting at the junction between two helices belongs to
the first one. We let $n_i = s_i - s_{i-1}$ denote the lenght of  helix $i$.
We define also 
\be
u_i = \sigma_i \frac{L}{a_i \sqrt{1+h_i^2}} \;\;\;,
\label{u}
\ee
where L is the lenght of a peptide unit,
so that  the line element on each helix is
$\left|\dot{{\bf h}}_i\right| ds = L ds$. We assume the sign
$\sigma_i = \pm 1$ of $u_i$
positive for right-handed and negative for left-handed helices,
while the product $u_i h_i$ is always positive.
We also ask that helices have the same lenght of the chain they
represent, setting $\Delta s=1$ for a peptide-unit  move 
along the protein chain. This requirement implies 
that $n_i$ as defined above coincides with the number of residues 
in the helix.

In order to write down a simple hamiltonian, we further simplify the
model, resorting to  the following variables:

\be
\begin{array}{c c l l}
N_h\!\!\!&~~ &\!\!\hbox{the total number of helices} \;\;\;&~~ \\
n_i\!\!\!&=&\!\!s_i - s_{i-1} &(n_i \in [p_1,p_2])\\
l_i\!\!\!\!&=&\!\!\frac{1}{2}\left(s_i + s_{i-1}+1 \right)&~~ \\
{\bf v}_i\!\!\!&=&\!\!{\bf h}_i(s_i) - {\bf h}_i(s_{i-1})&~~\\
{\bf B}_i\!\!\!&=&\!\!\frac{1}{2}({\bf h}_i(s_i) + {\bf h}_i(s_{i-1}))& 
~~ \\  
\end{array}
\label{dyva}
\ee
where $p_2 = N - (N_h-1) p_1$ and $p_1=3$, since a helix cannot be
defined with less than three residues. $n_i$ is
the lenght of the $i$-th helix  expressed in residues; $i \in [1, N_h]$;
 $l_i$ represents
the position along the sequence of the center of the $i$-th helix;
${\bf v}_i$ is
the vector joining the end-points of helix ${\bf h}_i$;
${\bf B}_i$ is the
the spatial position of the middle point of ${\bf v}_i$.

Two other variables  are necessary to
specify the "shape" of a helix: a particularly 
useful choice  is to
introduce:
\bq
z_i &=& \frac{L \tau_i}{u_i} \;\;\;, \\
\label{z}
w_i &=& u_i - 2 \pi \vartheta(-u_i)\;\;\;,
\label{w}
\eq
where 
$u_i = L \sigma_i (\kappa_i^2 + \tau_i^2)^\frac{1}{2} $
($\kappa_i$, $\tau_i$ are the constant curvature and torsion of the
$i$-th helix) and $\vartheta (\bullet )$ is the Heaviside function.
The definition of $w_i$, in $w_i \in [0, 2\pi]$,
allows us to remove the discontinuity between
right and
left-handed helices at $u=\pm \pi$, which is model-induced but
inevitable in a
description of the chain in term of helices.
The sequence enters the model through the variables $q_k$ ($k=1 \ldots N$) 
and ${\bf p}_{\perp}^2(l,w)$.
The former are related to the nature of each residue $k$, and measure  
its coupling to the other residues, 
due to the fact that the Mijazawa-Jernigan  
interaction matrix \cite{mije85} can be written~\cite{litaa95} as:
\be
M_{\rho \sigma} = \mu_0 + \mu_1 (q_\rho + q_\sigma) + \mu_2 q_\rho q_\sigma
\;\;\;\;\;\;\;\; (\rho,\sigma=1, \ldots ,20) \; .
\label{mijemx}
\ee

Since we deal with entire helices at a time, and not with single
residues, we introduce the average $q$ of a helix, centered
in $l_i=l$, as
\be
\overline{q}(l) = \cases{\frac{1}{2m + 1}\sum_{j=-m}^m q_{l+j},&if $l=1,2,\ldots$ \cr
\frac{1}{2(2m + 1)}\sum_{j=-m}^m (q_{l-\frac{1}{2}+j}+q_{l+\frac{1}{2}+j}),
&if $l=\frac{1}{2},\frac{3}{2},\ldots$}
\label{qbar}
\ee
(integer or half-integer values of $l$ are the only ones allowed for 
the central points of the helices, $l_i$;
the variable $m$ is an arbitrary number, 
comparable with the mean lenght of the helices).

The other variables are defined by:
\be
{\bf p}_{\perp}^2(l,w,n) = \frac{1}{(\sum_{j=-n}^n Q_{l+j})^2}
 \sum_{j,k=-n}^{n} Q_{l+j} Q_{l+k}  \cos((j-k) w) \;\;\;,
\label{pquad}
\ee 
where ${\bf p}_{\perp }(l,w,n)$
is the projection on the plane perpendicular to the helix axis 
of the "hydrophobic dipole moment", calculated at a point on the axis 
and normalized with respect to the total hydrophobic charge~\cite{litaa95}
$ \sum_{j=-n}^n Q_{l+j}$, where:
\(Q_\rho=\mu_0/2 + \mu_1 q_\rho  + (\mu_2/2) q_\rho^2 \). 

The quantity  ${\bf p}_{\perp }^2$
reveals the 
prevalence of non polar residues on one side of the helix, characterized
by the periodicity $w$.

\noindent
The following constraints hold among the variables previously defined:  
\begin{enumerate}
\item{the sum of the residues of all the helices must be equal to the
total lenght of the chain:
\be
\sum_{i=1}^{N_h} n_i - N =0  \; ; 
\label{vinc0}
\ee
}
\item{the lenght of $v_i$ is related to the
lenght and shape of the helix:
\be
{\bf v}_i^2 - \left| {\bf h}_i(s_i) -
{\bf h}_i(s_{i-1}) \right|^2  
\equiv  {\bf v}_i^2 - n_i^2 L^2 
\left[ z_i^2 + (1-z_i^2)
\frac{\sin^2(\theta_i)}{\theta_i^2}\right] = 0 \;\;\;, 
\label{vinc1}
\ee
where $\theta_i = n_i u_i / 2$;}
\item{the end of one helix must coincide with the beginning of the
following one, both in sequence and in space:
\bq
 {\bf B}_i - {\bf B}_{i-1} - \frac{({\bf v}_i + {\bf
v}_{i-1})}{2}&=& 0\;\;\;,
\label{vinc2} \\
l_i - l_{i-1} -
\frac{n_i + n_{i-1}}{2} \;\;&=& 0\;\;\;.
\label{vinc3}
\eq   
}
\end{enumerate}
In these equations, $i$ ranges from 1 to $N_h$, and, to be consistent
with the definitions of $l_i$, we set $l_0 = 1/2$, $n_0 = 0$. 

With the above defined
variables we write a hamiltonian of the form: 
\be
H = H_{nn} + \sum_{i=1}^{N_h} (H_i^0 + H_i^1) + \sum_{i<j=2}^{N_h}
H_{i,j}
\;\;,
\label{protham}
\ee
where we have defined:
\bq
H_{nn} &=& \gamma_2 (N_h-1) \;\;\;,\nonumber \\ H_i^0 &=& (n_i-1)
\gamma_0 \left[c_1 \left((w_i- c_2)^2 - c_3 \right)^2 + c_4 +c_5
\left(z_i-c_6 + c_7 (w_i-c_8)^2 \right)^2 \right] \;\;, \nonumber \\
H_i^1 &=& - \gamma_1 n_i P(l_i,w_i)\;\;\;,\nonumber \\
H_{ij} &=& \vartheta(\rho_1 - \Delta B_{ij}) \vartheta(\Delta B_{ij}-\rho_0) 
\left[ \gamma_3
\chi \left(\mu_0 + \mu_1 \left(\overline{q}(l_i) + \overline{q}(l_j)\right)
+ \mu_2  \overline{q}(l_i) \overline{q}(l_j) \right) \right] + \nonumber\\ 
&&+ \gamma_4 \vartheta(\rho_0- \Delta B_{ij})\; . \nonumber
\eq
Here $\gamma_i$ are dimensional parameters weighting the various
contributions, while $c_k$ are known adimensional constants and
$\Delta B_{ij} = \left| {\bf B}_i- {\bf B}_j \right|$.

Costraints will be implemented explicitly, by direct substitutions of
the variables in the above hamiltonian, which will eventually be 
written as a 
function  of the independent variables.

A detailed discussion of the various terms appearing in
Eq.~(\ref{protham}) has been given elsewhere~\cite{brus97}; here
we just recall that $H_i^0$  recovers  in an effective way the
the experimental
Ramachandran plot~\cite{rasa68}, 
thus dictating which kind of helices are more likely to be
formed. $H_i^1$, on the other hand, is sequence dependent and favours the
separation of polar and non polar residues on the helices:
$P(l_i,w_i) = {\cal F}({\bf p}_{\perp}^2(l_i,w_i,n))$
is some simple function of ${\bf p}_{\perp}^2(l_i,w_i,n)$.

$H_{nn}$ represents an extremely simplified way to keep next-neighbours
interactions into account: a constant, positive energy is involved in
helix breaking, independently on their orientation.
$H_{ij}$ has the
simple form of a square-well with an infinite barrier on one side,
representing hard core repulsion between helices. The interaction, in
the range $\Delta B_{ij} \in [\rho_0,\rho_1]$
has the form of Eq.~(\ref{mijemx}),
calculated with the average "charges" $\overline{q}_i$ of the helices.

For the sake of simplicity inter-helical hydrogen bonds
are not distinguished from hydrophobic
interactions (hence we disregard their dependence on orientation),
and both are described by $H_{ij}$. 

\section{Analysis of the four proteins}
\label{sec:an}

We now consider Eq.~(\ref{protham}) in the limit $\gamma_1 \ll \gamma_0$,
without non-local interactions
($\gamma_3=\gamma_4=0$) and at fixed number of helices $N_h$,
and ask ourselves to
what extent the correct native secondary structure can be recovered by
local terms only.

The former limit is equivalent to studying the ground state of 
$H^1 = \sum_{i=1}^{N_h} H_i^1$ with only two allowed values
$(w_\alpha, w_\beta)$
for each $w_i$, corresponding respectively to $\alpha$ and
$\beta$ configuration. We shall look for
the values of
$(n_i, w_i)$, at fixed $N_h$, which best  represent the native
secondary structure, in the cases of two synthetic sequences~\cite{kasca93}
and of two natural proteins, identified by PDB codes~\cite{pdb} 2mhr
(myohemerythrin) and 2asr (aspartate receptor, ligand binding domain).
These proteins are known to fold in the "four-helix bundle" conformation.

We assume that the function $P(l_i,w_i)$, appearing in the
expression of $H_i^1$, 
has the form:
\be
P(l_i,w_i) = \cases{ p_{\perp}^2(l_i,w_i,3), &if $l_i$ is an integer, \cr
\frac{1}{2}\left[p_{\perp}^2
(l_{i}-\frac{1}{2},w_i,3)+ p_{\perp}^2(l_{i}+\frac{1}{2},w_i,3)
\right], &if $l_i=k+\frac{1}{2}$, for integer $k$.}
\label{bigp}
\ee
We have chosen $n=3$ in expression (\ref{pquad}) since this involves
calculating the hydrophobic dipole on an helix of seven residues, a
reasonable lenght both for $\alpha$-helices and for
$\beta$-strands.

First of all we plot $P(l,w_\alpha)$,   
$P(l,w_\beta)$ for all the proteins:
Figures~(\ref{fig:seqB}, \ref{fig:seqF}, \ref{fig:2asr}, \ref{fig:2mhr})
reveal that indeed a clear dominance of 
$P(l,w_\alpha)$ seems to be a sufficient condition for
$\alpha$-helices, though not a necessary one.

Then we study the ground state of the local hamiltonian $H_1$:
we set $\gamma_1=1$ and
exhaustively search the configuration space with $N_h=4$, recording
the best ten configurations we find.
The choice of $N_h$ is suggested by our {\sl a-priori} knowledge
of the native state of these proteins, and by the reasonable assumption
that the
existence of short turns is related rather to the three-dimensional 
structure
than to sequence periodicity requirements, so that they could not be 
efficiently recovered by the local hamiltonian.
 
To test the goodness of the configurations we find, we proceed as
follows:
first of all we divide each protein into four parts, corresponding to
the four ``arms'' 
in the native bundle conformation, and look at those which are in a
helical configuration (for 2asr, we consider the short  $3_{10}$-helices
together with $\alpha$-helices). 

Then we consider our configurations and 
compare each element in the bundle with the corresponding native one,
and count the residues that have been correctly predicted as belonging
to an $\alpha$-helix. If $n_\alpha$ is their number, the quantities:
\be
C_{tot}=\frac{n_\alpha}{N}\;\;\; , \;\;\;C_{rel}= 
\frac{n_\alpha}{n_\alpha^{nat}} \; \; , 
\label{corr}
\ee
will give the percentage of success in relation respectively  
to the total number of
residues and to the number
$n_\alpha^{nat}$ of residues belonging to helices
in the native state. 

We obtain the following results:\\
\begin{center}
\begin{tabular}{l|c|c|c|c|c}
protein \hspace*{.2cm}&\hspace*{.2cm} energy \hspace*{.2cm}    
& \hspace*{.2cm}  $(n_1, n_2, n_3, n_4)$ \hspace*{.2cm} 
& \hspace*{.4cm} helix \hspace*{.4cm} &
\hspace*{.2cm}    $C_{tot}$ \hspace*{.2cm}&
\hspace*{.2cm} $C_{rel}$\\ \hline
seqB & -20.290 & (25, 3, 27, 29) & $(\alpha, \alpha, \beta, \alpha)$
&0.42& 0.55\\ 
& -20.170& (25, 13, 19, 17) & $(\alpha, \alpha, \alpha, \alpha)$
&0.70&0.93\\ \hline
seqF & -19.203 & (5, 39, 13, 17) & $(\alpha, \alpha, \alpha, \alpha)$
&0.55& 0.73\\ 
& -19.062& (11, 27, 19, 17) & $(\alpha, \alpha, \alpha, \alpha)$
&0.69&0.91\\ \hline
2asr & -36.550 & (3, 3, 129, 7) & $(\beta, \beta, \alpha, \alpha)$
&0.24& 0.27\\ 
& -35.830& (4, 3, 127, 8) & $(\beta, \alpha, \alpha, \alpha)$
&0.25&0.28\\ \hline
2mhr & -35.297 & (5, 47, 49, 17) & $(\beta, \alpha, \alpha, \beta)$
&0.24& 0.34\\ 
& -35.192& (39, 13, 49, 17) & $(\alpha, \alpha, \alpha, \beta)$
&0.42&0.60\\ \hline
\end{tabular}
\end{center}
For each protein the first line refers to the ground state, while the
second refers to the configuration with the highest correlation to the
native state, among the ten recorded. The most native-like 
conformations for the four proteins
appear at position 3, 8, 9, 2 respectively, in the list of the best
ten configurations.

For both the synthetic sequences  native
$\alpha$-helices correspond to residues (3-16; 22-35; 41-54; 60-73); the
secondary structure of 2mhr presents 
$\alpha$-helices at positions (12-14, 19-37; 41-64; 70-85; 93-109,
111-114); that of 2asr shows $\alpha$-helices 
at positions (2-38; 49-72; 80-104; 117-141),
while residues 44-48, 77-79 are in $3_{10}$ conformation.

\section{Comments and conclusions}
\label{sec:conc}
In this letter we addressed three questions:
how good is the hydrophobic dipole
moment in describing the relationship between sequence and
secondary structure? What is the role of local terms in the
hamiltonian? Is it possible to predict the native secondary structure 
on the grounds only of the hydrophobic dipole?

The results we obtain 
show that the choice of describing the sequence periodicity by means of
${\bf p}_\perp^2(l,w,n)$ and $P(l,w)$ (Eqs.(\ref{pquad},\ref{bigp}))
is substantially correct, 
both at a descriptive and at a more quantitative level.

Indeed, a qualitative
correlation is evident between regions where $P(l,w_\alpha)$
dominates and the position of $\alpha$ helices, in all the proteins
considered. It is however not straightforward to describe
this correlation quantitatively, since it is not easy to 
unambiguously express in mathematical language
what one should recognize as "dominant". For this reason, we cannot
exclude that better definitions than Eq.(\ref{bigp},\ref{pquad})
 may  be found to characterize local
periodicities in the sequence, even if we consider our choice to be a
reliable one. 

Moreover, we have introduced an objective way 
to assess how similar is the ground
state to the native one, and indeed the minimal energy configurations we find
suggest that our variables and hamiltonian are not so bad in
describing the system.

It can indeed be noticed  
that, despite the strong simplifications introduced in 
considering only $\alpha$ and $\beta$ helices and in taking $N_h=4$ (that
forbids a simultaneous 
description of both the helices and the turns), we obtain
good results for the two synthetic sequences: among the  
low energy states a configuration is found which shows a high degree
of correlation to the native secondary structure, 
and the fact that this configuration is not the ground state can be
considered a minor problem, at this level of simplification.

The results for 2mhr and 2asr, on the other hand, leave us with several 
open questions about the relative importance of local and nonlocal
terms in the hamiltonian. 
The hydrophobic moment diagrams Fig.(\ref{fig:2asr},
\ref{fig:2mhr}) are more complex than those for the synthetic proteins, 
which could signal a minor importance of the local terms with respect
to the nonlocal ones. Indeed, it is commonly  believed that
the secondary structure results from the need to maximize 
compactness of the protein 
and protection of the non-polar residues from water. 
According to these ideas
the periodicity of the sequence could be an outcome of evolution,
useful to remove a possible  source of
frustration and prevent misfolding, while 
increasing the stability of the native state; 
yet  proteins need
not be optimized with respect to their periodicity.     

On the other hand, the poor results we obtain with these proteins could
also be  partially due to the approximations introduced, and we cannot
exclude that better predictions could be obtained  just resorting to 
a more complete expression of the local  terms.
A more definite answer to the above questions is left to future efforts.


\begin{figure}
\caption{Plot of $P(l,w_\alpha)$ (continuous line) and $P(l,w_\beta)$ 
(dotted line) for the sequence seqB.}
\label{fig:seqB}
\end{figure}

\begin{figure}
\caption{Plot of $P(l,w_\alpha)$ (continuous line) and $P(l,w_\beta)$ 
(dotted line) for the sequence seqF.}
\label{fig:seqF}
\end{figure}

\begin{figure}
\caption{Plot of $P(l,w_\alpha)$ (continuous line) and $P(l,w_\beta)$ 
(dotted line) for the protein 2asr.}
\label{fig:2asr}
\end{figure}

\begin{figure}
\caption{Plot of $P(l,w_\alpha)$ (continuous line) and $P(l,w_\beta)$ 
(dotted line) for the protein 2mhr.}
\label{fig:2mhr}
\end{figure}

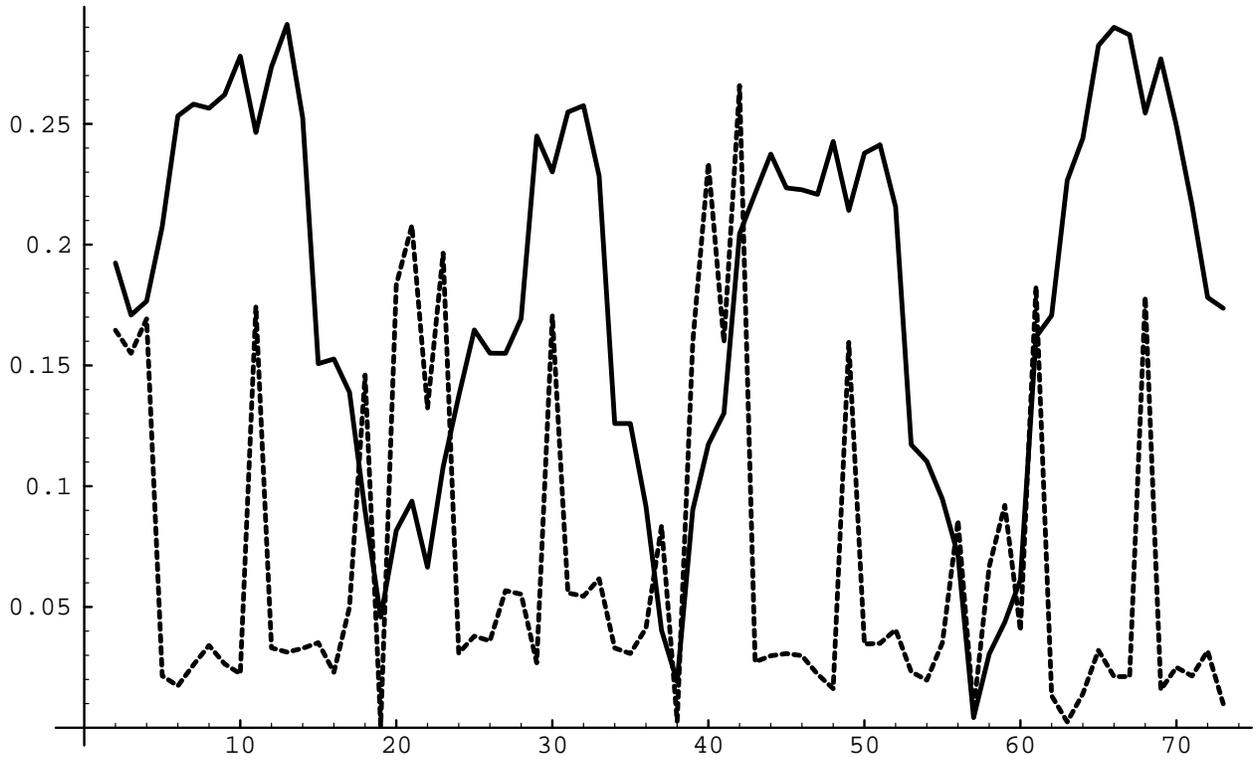
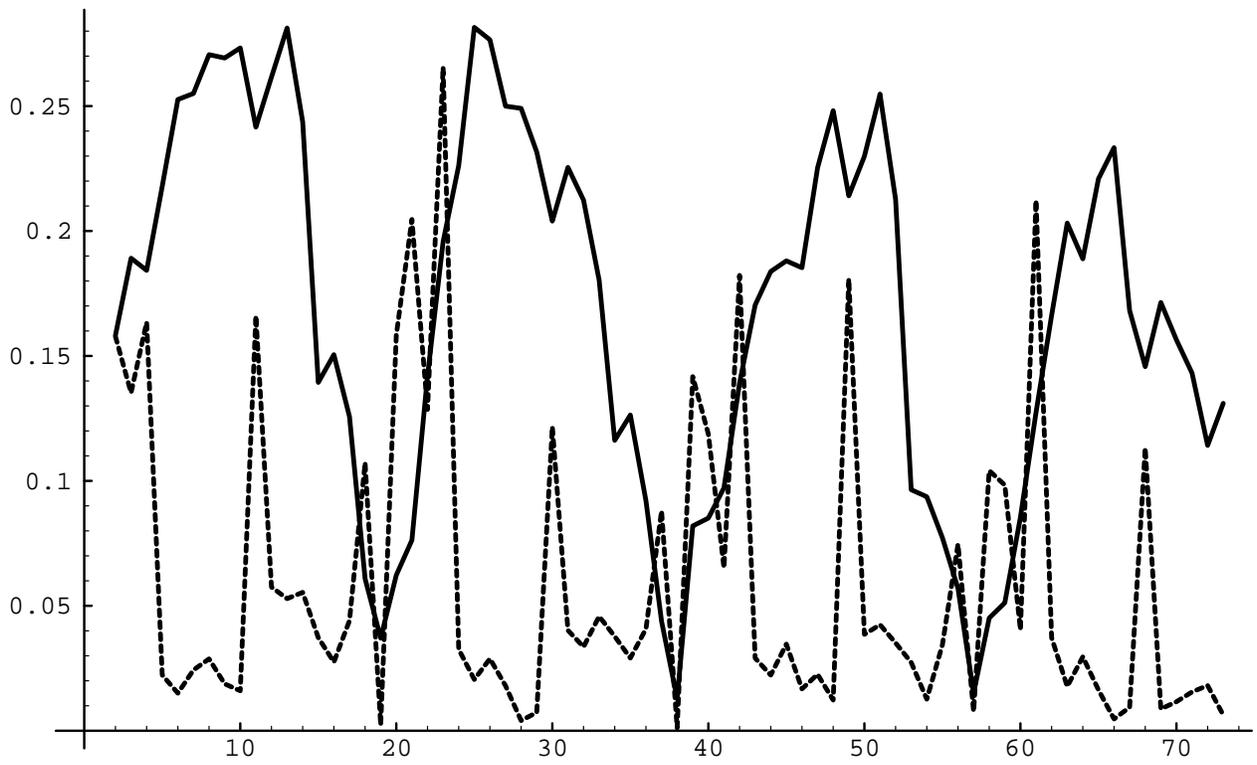
\end{document}